\documentclass[11pt,reqno]{amsproc}
\allowdisplaybreaks
\usepackage{cite}
\usepackage{tcolorbox}
\tcbuselibrary{breakable}
\usepackage{enumerate}
\usepackage{psfrag}
\usepackage{caption}
\usepackage{color}
\usepackage{fullpage}
\usepackage{graphicx}
\usepackage{mathrsfs}
\usepackage{subfigure}
\DeclareGraphicsExtensions{.eps}
\usepackage{ucs}
\usepackage[utf8x]{inputenc}
\usepackage{epstopdf}
\usepackage[semicolon,square,authoryear]{natbib}
\newlength{\drop}
\definecolor{amethyst}{rgb}{0.6, 0.4, 0.8}
\definecolor{burgundy}{rgb}{0.5, 0.0, 0.13}
\title{\textbf{Bioremediation of oil and heavy metal contaminated soil in construction sites: a case study of using bioventing-biosparging and phytoextraction techniques}}

\author{\textbf{E.~Eslami$^{1 *}$}, and \textbf{S.~H.~S.~Joodat$^{2}$} \\
  {\small 1 Department of Earth \& Atmospheric Sciences, University of Houston. \\
  	\small 2 Department of Civil \& Environmental Engineering, University of Houston.\\
    \textbf{* Correspondence to:}~\textsf{eeslami@central.uh.edu}}}

\keywords{bioremedation; bioventing; biosparging; phytoextraction; plant-assistant remediation; cement pollution; concrete plant}

\begin{document}

\date{\today}

\begin{titlepage}
  \drop=0.1\textheight
  \centering
  \vspace*{\baselineskip}
  \rule{\textwidth}{1.6pt}\vspace*{-\baselineskip}\vspace*{2pt}
  \rule{\textwidth}{0.4pt}\\[\baselineskip]
       {\Large \textbf{\color{burgundy}
           Bioremediation of oil and heavy metal contaminated soil in construction sites: a case study of using bioventing-biosparging and phytoextraction techniques}}\\[0.3\baselineskip]
       \rule{\textwidth}{0.4pt}\vspace*{-\baselineskip}\vspace{3.2pt}
       \rule{\textwidth}{1.6pt}\\[\baselineskip]
       \scshape
       \vspace*{1\baselineskip}
       Authored by \\[\baselineskip]

   {\Large E.Eslami\par}
   {\itshape Graduate Student \\
   	Department of Earth \& Atmospheric Sciences \\
  	University of Houston, Houston, Texas 77004--5007 \\ 
  	\textbf{phone:} +1-832-205-5004, \textbf{e-mail:} eeslami@central.uh.edu}\\[0.75\baselineskip]
            
  {\Large S.~H.~S.~Joodat\par}
  {\itshape Graduate Student \\
  Department of Civil \& Environmental Engineering \\
  University of Houston, Houston, Texas 77204--4003}
  \vspace{0.1in} 

\end{titlepage}

\begin{abstract}
  This paper studies the bioremediation techniques utilized for degrading soil pollutants in a construction site located in Iran. The activities in a construction site, located in Garmdareh, Karaj, polluted the soil environment with a large amount of contaminants among which BTEX, PAHs and heavy metals are considered to be the most important components. 
  Employing conventional physicochemical techniques for remediation of such a polluted soil is both technically and economically challenging. Bioremediation, which involves taking advantage of the microorganisms for the removal of pollutants, is the most promising technology which is both relatively efficient and cost-effective. In this study, the coupled bioventing-biosparging technique and the phytoextraction process were utilized to remediate the deep and shallow layers of the polluted environment, respectively. 
  However, the widespread presence of the cement components in the site influenced the performance of bioremediation techniques. The results of this study indicated that the cement-polluted soil lowered the degradation rate of both BTEX (Ethylbenzene) and PAH (Pyrene) pollutants within deep layers. 
  Conversely, the degradation rate of Pb was increased in phytoextraction process due to proper stabilization of the surficial soil in presence of cement components. This study provides a critical view on the limitations and influential parameters of function-directed soil remediation techniques dealing with the presence of cementitious materials in construction sites.
   \end{abstract}

\maketitle

\vspace{-0.4in}

\section{INTRODUCTION}
Nowadays, the number of sources contributing to the environmental pollution is growing which highlights the need for taking advantage of the existing remediation approaches as well as developing new methods for its treatment \citep{khodaverdi2008fuzzy,eslami_Alizadeh_Omidifard_2013}. Among the various types of environmental issues that threat human's well-being everyday, soil contamination is considered as one of the most important problems of our time which is commonly originated from the construction and concrete producing sites all over the world \citep{pant2016bioremediation,lu2015impacts,cachada2018soil}. Different types of contamination after being released from their sources, can propagate throughout the heterogeneous medium of the soil via the fluid flow based on their initial concentration, the flow rate and the medium properties such as the permeability and can pollute the shallow and deep layers of the soil \citep{Nakshatrala_Joodat_Ballarini_P2,joshaghani_Joodat_Nakshatrala_2018,Nakshatrala_Joodat_Ballarini_a_2018}. Emission sources in such sites, that can increase the soil contamination level up to a harmful range, include but are not limited to the activities of heavy vehicle equipment, wastes produced during the handling processes, waste generated during the manufacturing processes, and construction waste disposal \citep{EPAExsituinsitu2006,krishna2007soil}. The most common soil contaminants released by aforementioned sources are the semi-volatile organic compounds (SVOCs) such as petroleum hydrocarbons (PHs) (e.g., fuels and oils) and polycyclic aromatic hydrocarbons (PAHs), BTEX (e.g., Benzene, Toluene, Ethylbenzene, Xylenes), incomplete-hydrated cements and cementitious components, as well as heavy metals such as Co, Cr, Hg, Ni, and Pb \citep{cai2008status,chen2015bioremediation,krishna2007soil}. High concentrations of these contaminants can cause harmful effects on the human health as well as the environment. Several studies have been performed worldwide to address the extent and severity of soil contamination and its environmental consequences \citep{khan2004overview,wilcke1998urban,wang2003soil,dong2001instances,tiller1992urban,krishna2007soil,khan2017soil,carre2017soil}. Therefore, it is essential to investigate reliable techniques and implement them in the field to attenuate the possible soil contamination. 

The drawbacks associated with the effective conventional methods, such as cost and complexity, are the key reason for pursuing alternative methods for the sufficient remediation of soil
contaminants \citep{dixon1996bioremediation}. Among the possible techniques for remediation of the polluted sites, bioremediation techniques are known as effective and innovative solutions for pollution reduction. These eco-friendly methods have been receiving growing attention, particularly due to their relatively high safety and low running cost compared to the physical and chemical methods, such as burning and adsorption \citep{megharaj2011bioremediation,shiomi2013novel,adams2015bioremediation,pant2016bioremediation,strong2008treatment}.
 These techniques provide the possibility to destroy or render the adverse effects of various contaminants using natural biological activities \citep{vidali2001bioremediation}. In addition, these techniques have been highly accepted by public and are able to be set up on the site \citep{natsheh2013multivariate}.

Bioremediation techniques can be classified into two categories depending on the location in which the treatment is implemented: in situ and ex situ. The in situ bioremediation can be described as the process whereby organic pollutants are biologically degraded, under the natural conditions, to either carbon dioxide, water or an attenuated transformation product \citep{EPAExsituinsitu2006,megharaj2011bioremediation}. The in situ techniques are generally inexpensive and need low maintenance compared to the ex situ ones which require the excavation of the contaminated samples for treatment \citep{khan2004overview}. Examples of in situ bioremediation techniques include bioventing, airsparging and phytoremediation, and common ex situ techniques include land treatment, composting, and biopiling \citep{EPAExsituinsitu2006}. 

The bioremediation techniques for reducing PHs, PAHs and other petroleum-related soil contaminants have been the topic of various research studies summarized in 
\citep{EPAExsituinsitu2006,ferrarese2008remediation,koshlaf2017soil}. Both in situ and ex situ techniques can be utilized for remediate contamination efficiently. Among such techniques, bioventing and biosparging have been proven to be effective for removing PH and PAH contamination \citep{erdogan2011bioremediation}. They can be easily applied to large areas and degrade the pollutants completely \citep{sayara2011bioremediation,mohan2006bioremediation}. The limitation, though, is that the bioventing-biosparging technique is only effective in deep layers of the polluted soil environment \citep{EPAExsituinsitu2006,boopathy2000factors}. For the shallow layers of soil, some studies have applied plant-assisted methods or phytoremediation, such as phytoextraction and rhizodegradation, to remediate the polluted soil especially when SVOCs and heavy metal pollutants are concerned. This method may require a long-term commitment and might be unable to remove the pollutants completely. Moreover, this method is highly cost-effective and environment-friendly compared with the other available approaches \citep{shukla2010bioremediation}.

This study has been performed on a construction site near Tehran used for producing precast concrete elements. The effects of cement as a soil contaminant along with petroleum-related components and heavy metals pollutants are studied on the efficiency of the bioremediation method. A combined bioventing-biosparging technique has been employed to remediate the deep soil layers in this construction site. Also phytoextraction as a phytoremediation technique has been utilized for removing the main heavy metal pollutants in cementitious soil.

\section{CASE STUDY}
The construction site is located in Garmdareh, near Karaj and 18 km west of Tehran and is elevated around 1304 meters above the sea level. The site’s total area is more than 6000 square meters and the contaminated area concerned in this study is around 2300 square meters (near one-third of the total area). The surrounding area of the site consists of both commercial and agricultural zones. Commercial zones include two refinaries, construction plants, manufacturing factories, etc. Agricultural zone include both open farms and green houses.

The main soil type of the site can be described as silty sand (SM) according to unified soil classification system with two locations. In the first location, a concrete plant used to operate for three continuous years. This site was mostly contaminated by petroleum related components such as diesel fuel, engine oil, etc., as well as cement products. Texture analysis demonstrated that the soil consisted of 81.8$\%$ sand/gravel, 11.1$\%$ silt and 7.1$\%$ clay. In the second location, a stainless painting process was performed for nearly one year which was later moved to an indoor area. This location was contaminated by heavy metals especially Pb, Cr and Co, as well as the cementitious components. The texture analysis of the soil showed a composition consisting of 75.5$\%$ sand/gravel, 13.6$\%$ silt, and 10.9$\%$ clay. The average ground water level was 67.9 meters and the average moisture content and pH level of soil were 32$\%$ and 7.9, respectively. Throughout the year, the temperature of soil at 3 m depth was between 22 to 41°C. The soil consisted of approximately 14$\%$ air-filled pores (at 3 meter depth) with oxygen saturation at a minimum level of 0.34$\%$. 

Several factors were taken into account for the selection of a proper soil bioremediation technique in this construction site. \emph{First}, the polluted area was wast and required the usage of techniques suitable for large-scale applications. \emph{Second}, the remediation cost should have been low to be financially feasible. \emph{Third}, the remediation process should have been nonstop since the activities causing the soil pollutants were performed continuously. \emph{Fourth}, the polluted area could have not been excavated because the facilities were already present and their movement could disrupt site’s activities. \emph{Fifth}, both shallow and deep layers of the soil have been contaminated which required a suitable combination of techniques for the soil to be fully recovered. \emph{Last but not least}, the sandy structure of the soil absorbed a huge amount of water due to its relative low moisture content. This resulted in a massive absorption gradient and semi-saturated soil condition which could reduce the efficiency of both bioventing and biosparging techniques \citep{EPAExsituinsitu2006}. Therefore, an exclusive bioremediation technique was needed to overcome the problems for the reliable removal of the pollutants. 

\begin{figure}
	\includegraphics[clip, scale=0.6,angle=270]{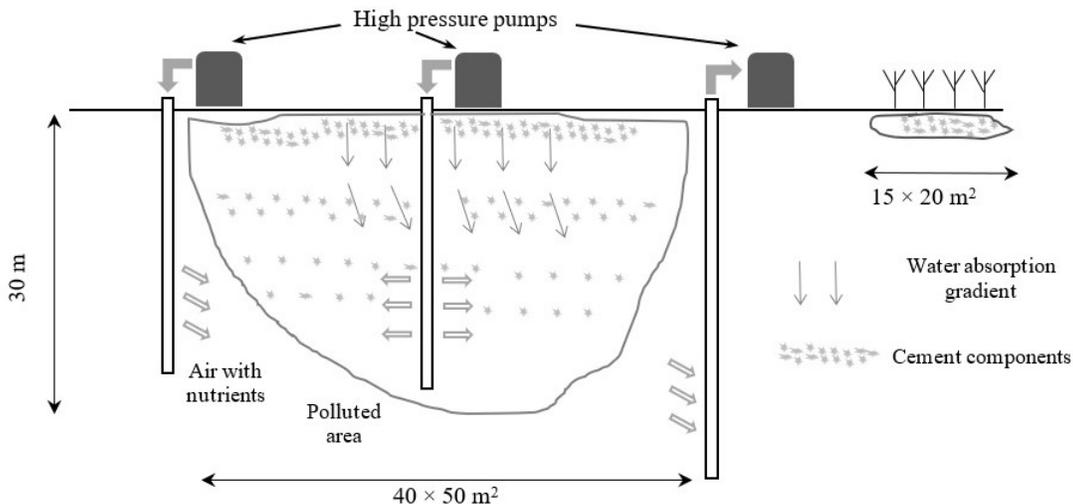}
		\vspace{-3cm}
	\caption{The schematic of the construction site: the bioventing-biosparging setup on the left and phytoextraction setup on the right.}
	\label{Fig1_Domain}
\end{figure}

Figure \ref{Fig1_Domain} shows a schematic of the construction site as well as the methodologies used for the remediation of polluted areas. In this study, a coupled technique using both bioventing and biosparging was employed to treat the deep layers of soil polluted by  petroleum related contaminants and cement related components. Treatment of shallow layers of the polluted soil was performed using a phytoextraction technique in which \emph{Brassica juncea} (or mustard green) was used to treat heavy metal contaminations such as Pb, Cr and Co.

\section{BIOREMEDIATION TECHNIQUES}
\subsection{Bioventing-biosparging for deep layers of soil}
Bioventing has been proven to be a useful technology for remediation of various sites under a variety of conditions. However, bioventing has some limitations \citep{boopathy2000factors}. One limitation is related to the ability to deliver oxygen to the contaminated soil. For example, soils with extremely high moisture content may be difficult to biovent because of the reduced soil permeability. Additionally, sites with shallow contamination levels can pose a challenge for the bioventing technique since designing a system that can minimize the environmental releases and can achieve a sufficient aeration might be difficult \citep{EPAExsituinsitu2006,azubuike2016bioremediation}. 

Biosparging involves the high-pressure injection of a gas (usually air or oxygen) and occasionally gaseous-phase nutrients into the saturated zones to promote aerobic biodegradation \citep{EPAExsituinsitu2006}. In air sparging, volatile contaminants (such as SVOCs) can also be removed from the saturated zones by desorption and volatilization into the air stream. Typically, biosparging is achieved by injecting air into a contaminated subsurface formation through a specially designed series of injection wells. The air creates an inverted cone of partially aerated soils surrounding the injection point. The air displaces pore water, volatilizes contaminants, and exits the saturated zone into the unsaturated zone. While in contact with ground water, oxygen dissolution from the air into the ground water is facilitated and supports the aerobic biodegradation \citep{EPAExsituinsitu2006,shukla2010bioremediation,azubuike2016bioremediation}.

\begin{figure}
\includegraphics[clip, scale=0.6]{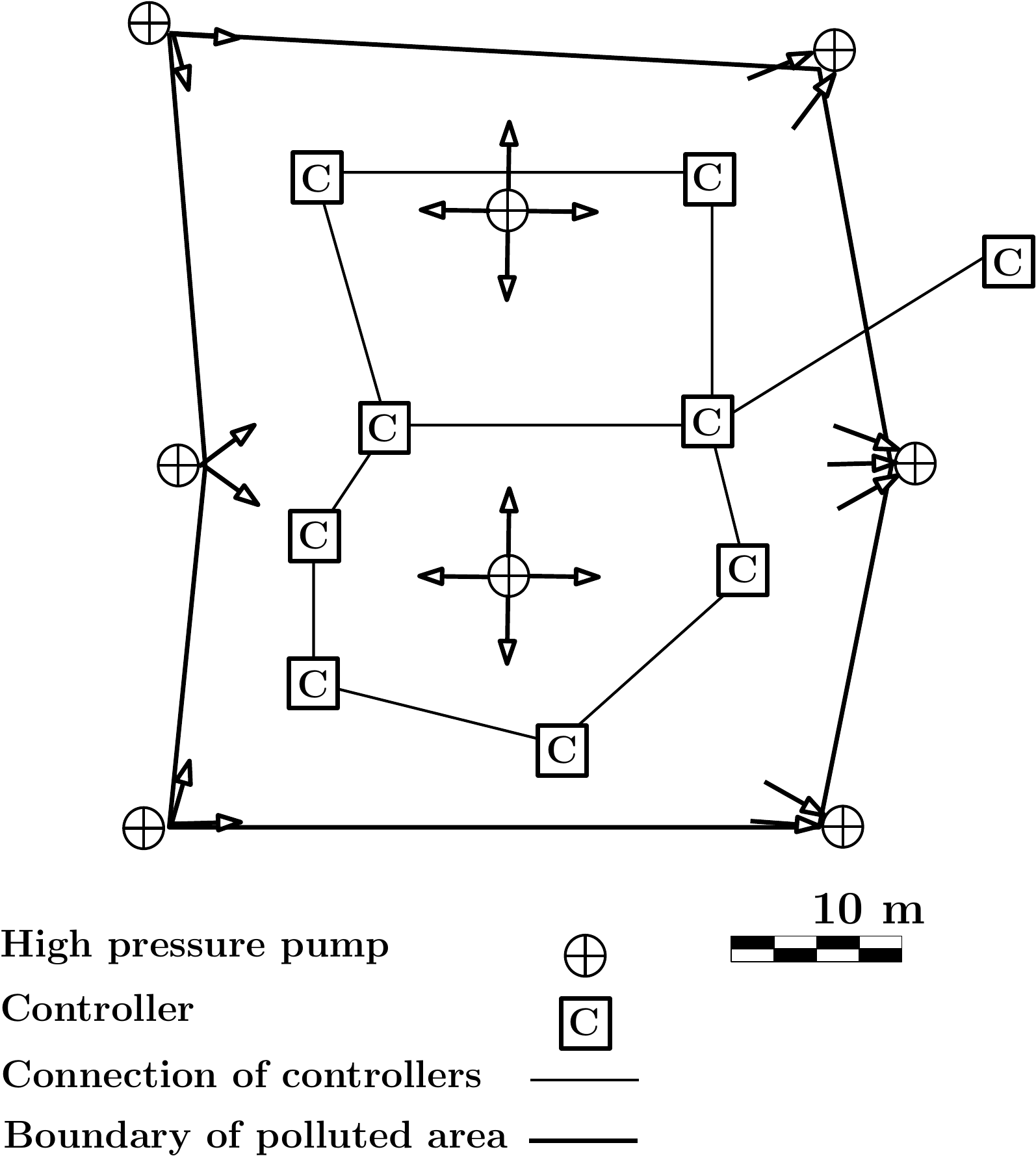}
\caption{The schematic bioventing-biosparging setup performed in the construction site.}
\label{Fig2_Schem_bio}
\end{figure}

As mentioned earlier, there is a downward gradient of water absorption in the soil (Figure \ref{Fig1_Domain}). This gradient was caused by the daily activities of the concrete plant, covering more than 2000 square meters of the site, in which the average of 150 liters of wastewater per hour was in contact with soil. Considering the soil condition along with the pollution sources of the equipment and the cement components, this volume of wastewater caused a semi-saturated polluted soil area as deep as 30 meters. As the semi-saturated soil condition can decrease the efficiency of bioremediation methods, both bioventing and biosparging techniques were utilized simultaneously to guarantee the efficient removal of the soil contaminants despite the unfavorable conditions. The coupled technique and its setup are shown in Figure \ref{Fig2_Schem_bio}. As can be seen in Figure \ref{Fig2_Schem_bio}, six high-pressure pumps were installed at the surroundings of the polluted area. Three of these pumps were used to inject the nutrient-rich air into the soil and the other three were installed to take out the air from the soil volume (bioventing technique).  In addition, two high-pressure pumps were installed inside the polluted area for injection of the oxygen-rich air into the semi-saturated soil (biosparging technique). The spacing of the pumps was around 25 meters. It should be noted that the pumps used for the bioventing process were alligned with the water absorption gradient to cover the maximum range of nourishing (see Figure \ref{Fig1_Domain}). Moreover, as shown in Figure \ref{Fig2_Schem_bio}, nine controllers were installed to monitor the contamination level of the soil, eight of which were monitored inside the polluted area.

\subsection{Phytoremediation for shallow layers of soil}
Plant-assisted bioremediation, or phytoremediation, is commonly defined as using green or higher terrestrial plants for treating chemically or radioactively polluted soils. Phytoextraction (or phytoaccumulation) uses plants or algae to remove contaminants from soils, sediments or water and convert them into harvestable plant biomass. Phytoextraction has been receiving growing popularity worldwide for the last twenty years or so \citep{shukla2010bioremediation}. Generally, this process has been employed more often for extracting heavy metals than for organics. At the time of disposal, contaminants are typically concentrated in the much smaller volume of the plant matter than in the initially contaminated soil. The main advantage of phytoextraction is that this procedure is an environmental friendly and cost efficient approach through which the heavy metals can be cleaned up from the soil without causing any kind of harm to the soil quality and with lower cost compared with any other process. Being controlled by the growth rate of the plant, the remediation process takes longer compared with the traditional soil cleanup processes. Also, the technique is proven to be effective mostly for shallow layers of the polluted soil (less than 60 cm in depth) \citep{cao2008trichoderma}.

In this construction site, the critical components causing contamination in the shallow layers of the soil included heavy metals such as Pb, Cr and Co caused by stainless painting processes. The average depth of the contaminated soil, expanded in approximately 300 square meters, was measured about 40 cm. The relatively low depth of the contamination can be attributed to two major facts. \emph{First}, the paining processes were performed temporarily and the processes had been moved to an indoor plant after less than a year. \emph{Second}, the existence of the cement components, caused either by site’s daily activities or the landscaping processes, resulted in the stabilization of soil’s surface area and consequently, the prevention of absorbance of heavy metals. Considering the target contaminants as well as their favorable depth, Brassica juncea was chosen for bioremediation. Brassica juncea has been extensively used by researchers to remove heavy metals especially Pb, Cr and Co. It has been proven that Brassica juncea can remove the contamination after two periods of harvesting in suitable soil and weather conditions \citep{shukla2010bioremediation,cao2008trichoderma}.

\subsection{Sampling}
In this study, monitoring of the contamination was carried out in the depths with the highest pollution level in the soil. These monitoring depths were 7 meters and 35 cm for bioventing-biosparging and phytoextraction techniques, respectively. As the soil is naturally able to remediate itself over time (self-remediation), this phenomenon should be taken into account for the analysis of remediation techniques. So, in this study, buckets (110 cm $\times$ 110 cm $\times$ 75 cm) of polluted soils were sampled at the beginning of the process to examine the self-remediation of both soils. The soil samples under either biological or physical conditions, were maintained in the lab for such an examination. Both samples and in situ techniques were monitored continuously for a duration of 50 days.

\section{RESULTS AND DISCUSSION}
\begin{figure}[!h]
	\subfigure[Ethylbenzene \label{Fig:3a_Ethylbenzene}]{
		\includegraphics[clip, scale=0.6]{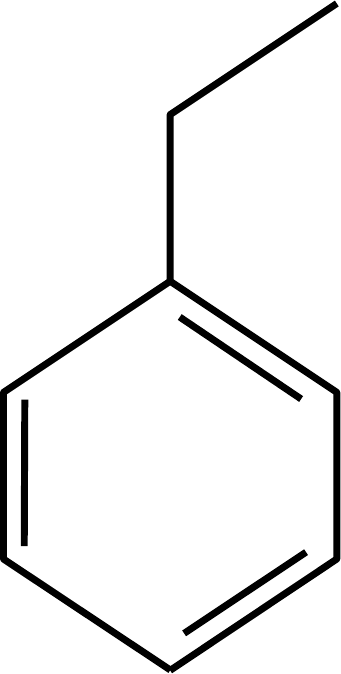}}
	\hspace{2cm}
	\subfigure[Pyrene \label{Fig:3b_Pyrene}]{
		\includegraphics[clip, scale=0.6]{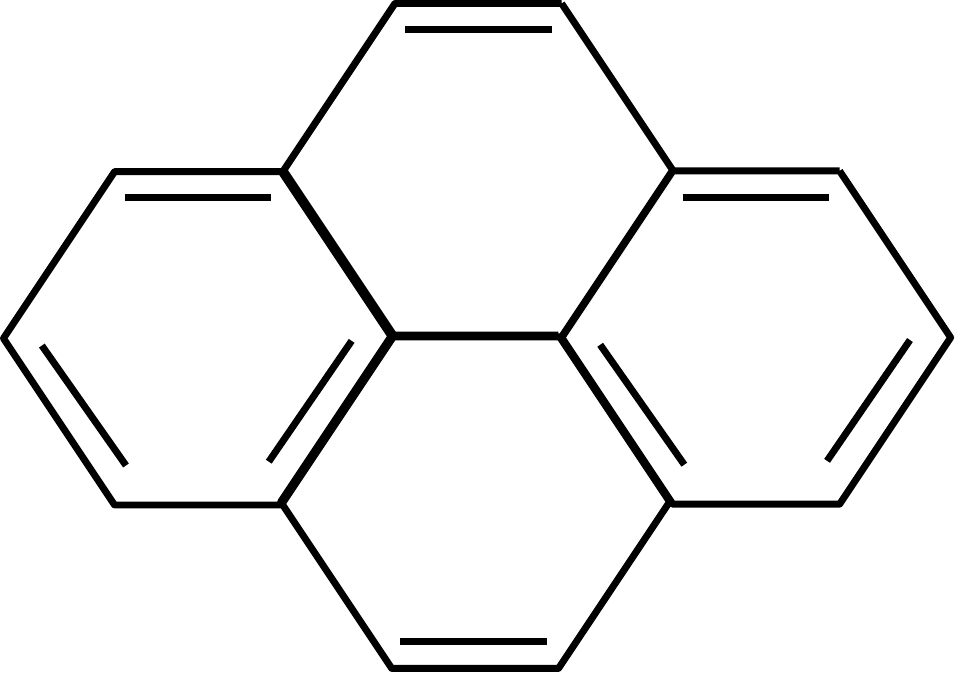}}
	\hspace{5cm}
	\subfigure[Lead (II, IV) oxide \label{Fig:3c_Lead}]{
		\includegraphics[clip, scale=0.6]{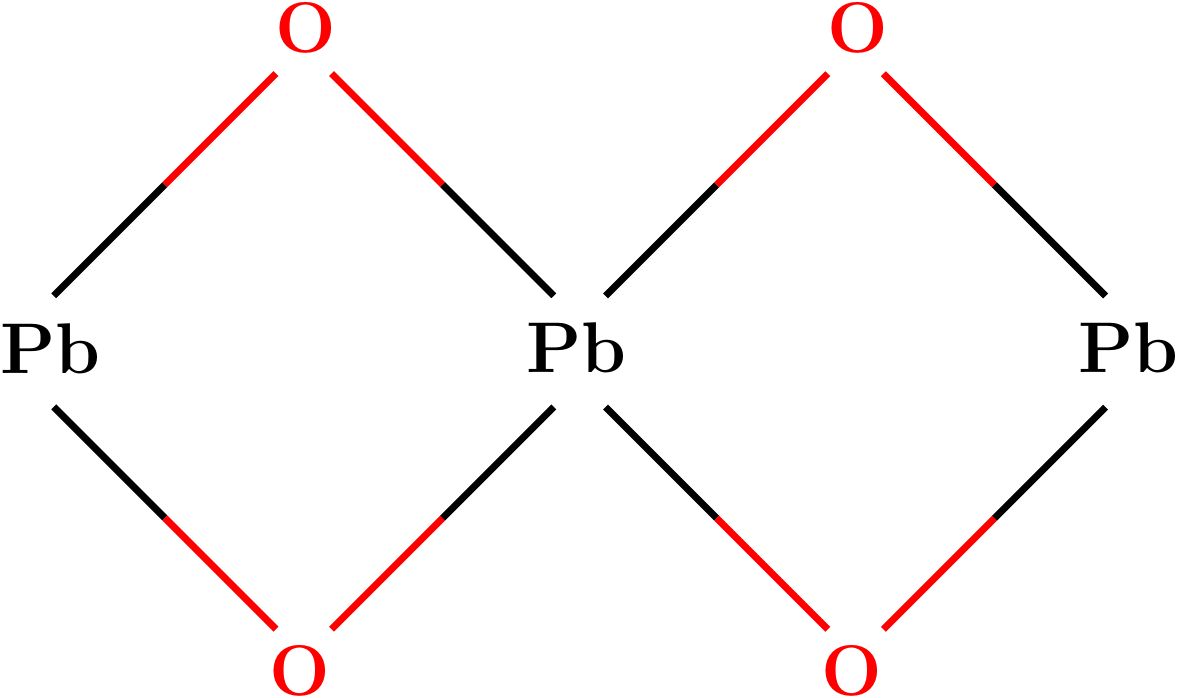}}
	\caption{Soil pollutants that have been treated using bioventing-biosparging techniques and phytoextraction. \label{Fig:Soil_pollutants_structure}}
\end{figure}

The soil pollutants considered in this study and their structural formulas are shown in Figure \ref{Fig:Soil_pollutants_structure} to illustrate the molecular interconnections. In addition, the results of the pollution level of Ethylbenzene and Pyrene after degradation are shown in Figures \ref{Fig:Ethylbanzene_Deg} and \ref{Fig:Pyrene_Deg}. These results represent the performance of the combined system of bioventing-biosparging technique in the deep layers of the soil in dealing with the BTEX (Ethylbenzene) and PAH (Pyrene) contaminations. Moreover, the results for the pollution level of Pb (as a heavy metal in form of oxide) after degradation at the shallow layers of the soil using phytoextraction are shown in Figure \ref{Fig:Pb_Deg}. 

\subsection{Bioventing-bisparging technique}
\begin{figure}[!h]
	\includegraphics[clip,scale=0.5]{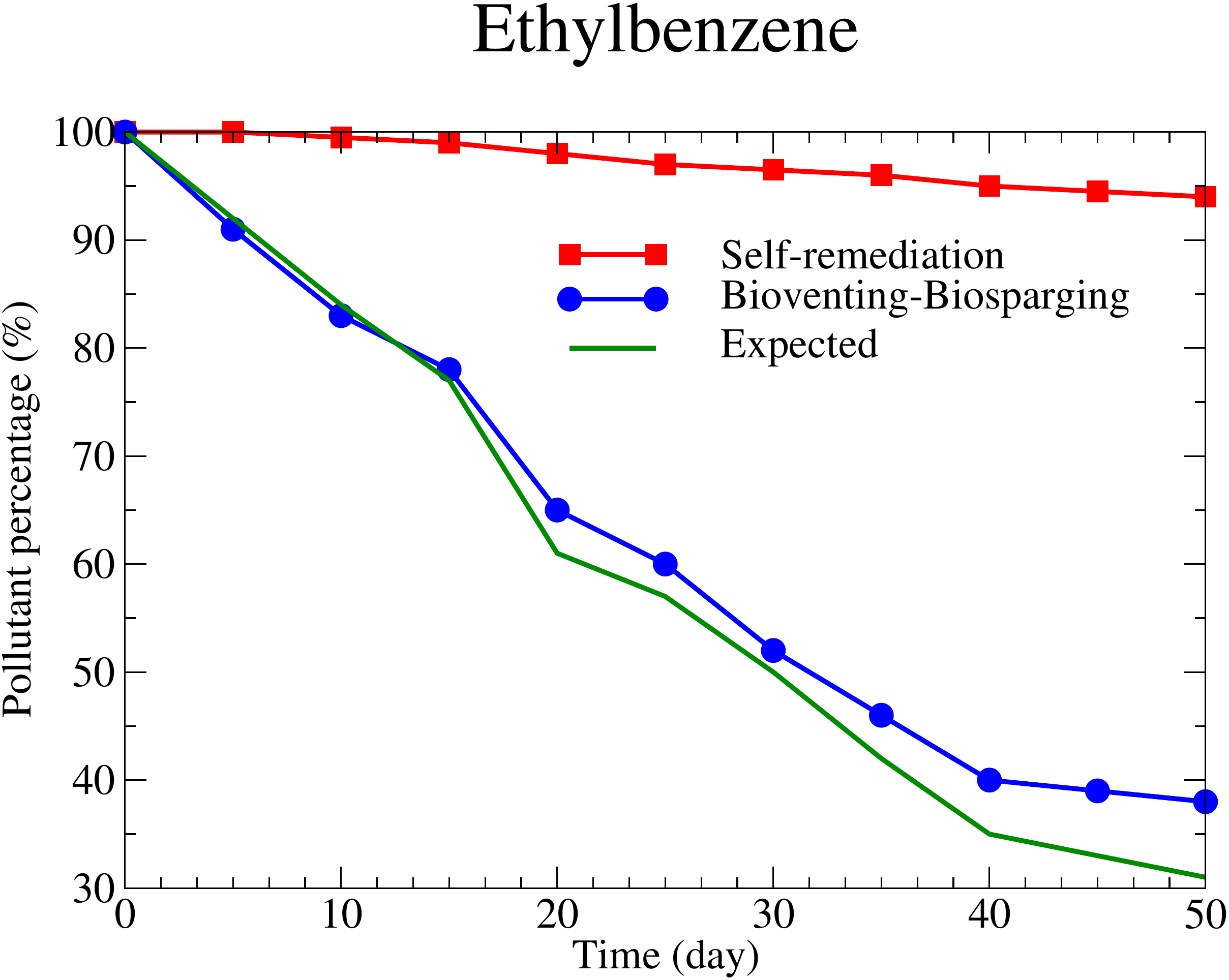}
	\caption{The remaining percentage of Ethylbenze after degradation in the deep layer using bioventing-biosparching technique in 50 days.}
	\label{Fig:Ethylbanzene_Deg}
\end{figure}

As can be seen in Figure \ref{Fig:Ethylbanzene_Deg}, the degradation rate of Ethylbenzene reaches its maximum level between days 15 and 40. The lower rates prior to day 15 can be attributed to the fact that the soil medium was getting prepared by injecting nutrient-rich air and it took 15 days for the soil to approach the most appropriate level of nourishment for advancing the remediation process. The reason for the decreasing rate of degradation observed after day 40 is bi-fold. First, the depth of the maximum concentration of Ethylbenzene has been changed because of the remediation process. Second, the level of concentration decreased over tine which lowers the degradation rate. Also the concentration of Etylbenzene at depth of 7 meters had decreased to around 60$\%$ by day 40. Therefore, in order to maintain a proper rate of degradation, it was needed to modify the parameters of the remediation process such as air pressures and the inserted nutrient concentration.

\begin{figure}[!h]
	\includegraphics[clip,scale=0.5]{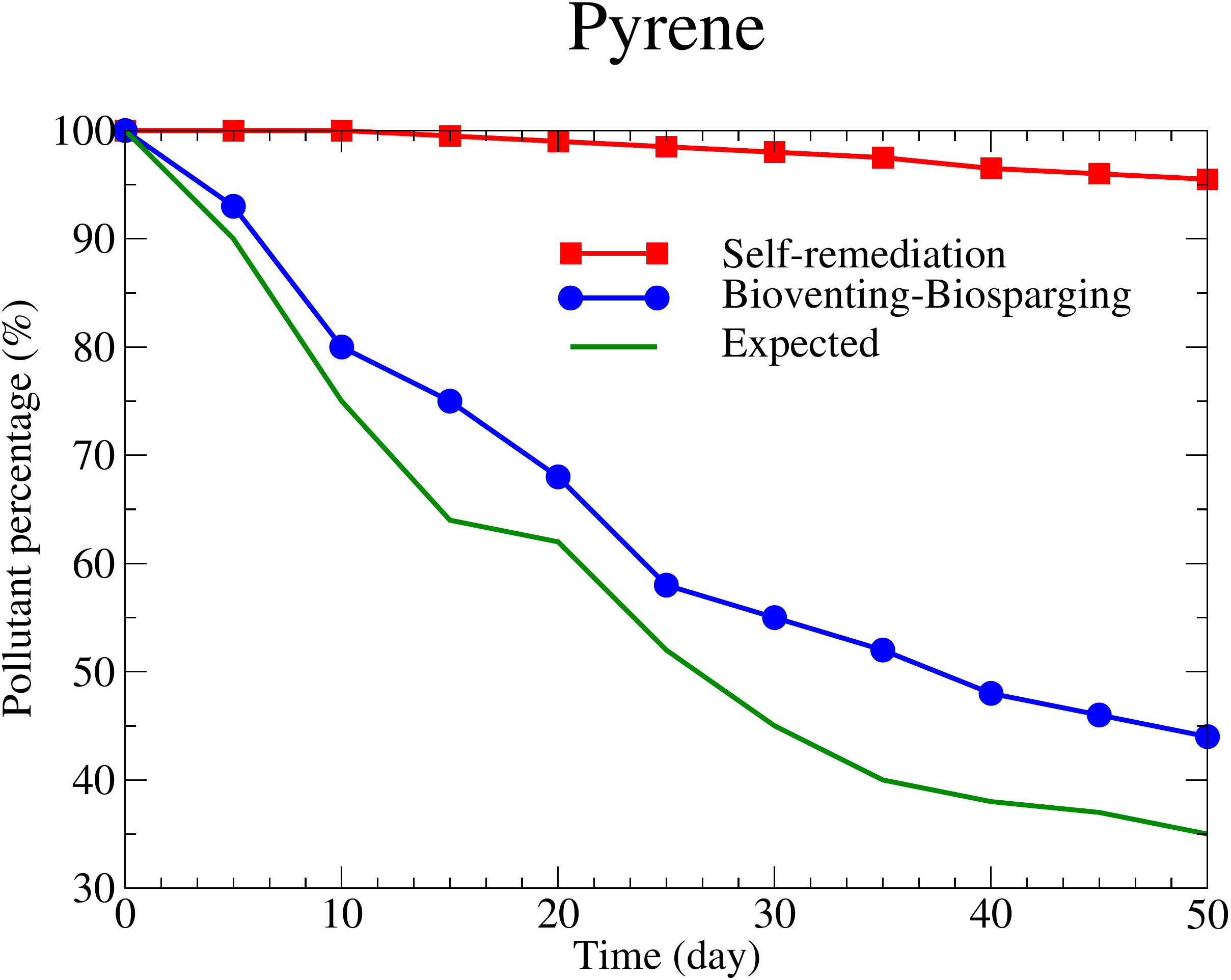}
	\caption{The remaining percentage of Pyrene after degradation in the deep layer using bioventing-biosparching technique in 50 days.}
	\label{Fig:Pyrene_Deg}
\end{figure}

Regarding Pyrene, according to Figure \ref{Fig:Pyrene_Deg}, the bioventing-biosparging technique had degraded the contaminants prior to day 25 at a higher rate compared with afterwards. Unlike Ethylbenzene, the degradation process had occurred from the beginning. However, similar to Ethylbenzene, the rate of degradation had decreased over time and the techniques should had been modified for obtaining more reliable results.

As can be concluded from Figures \ref{Fig:Ethylbanzene_Deg} and \ref{Fig:Pyrene_Deg}, the rate of degradation of Ethylbenzene was more than Pyrene. This difference in the rate of degradation can be attributed to their difference in the molecular structure. As a pollutant in BTEX family, Ethylbenezen’s structure consists of one cyclic hydrocarbon element (Figure \ref{Fig:3a_Ethylbenzene}). In contrast, Pyrene, like other PAHs, consists of multiple cyclic hydrocarbon elements (Figure \ref{Fig:3b_Pyrene}) which makes the degradation process more demanding. The other reason for the lower degradation rate of Pyrene compared to Ethylbenzene is the ability of soil in self-remediation towards each of these pollutants. As can be seen in Figures \ref{Fig:Ethylbanzene_Deg} and \ref{Fig:Pyrene_Deg}, the soil contaminated with Ethylbenzene demonstrates a better rate of self-remediation compared with the one contaminated with Pyrene. 

\subsection{Phytoextraction technique}
\begin{figure}[!h]
	\includegraphics[clip,scale=0.5]{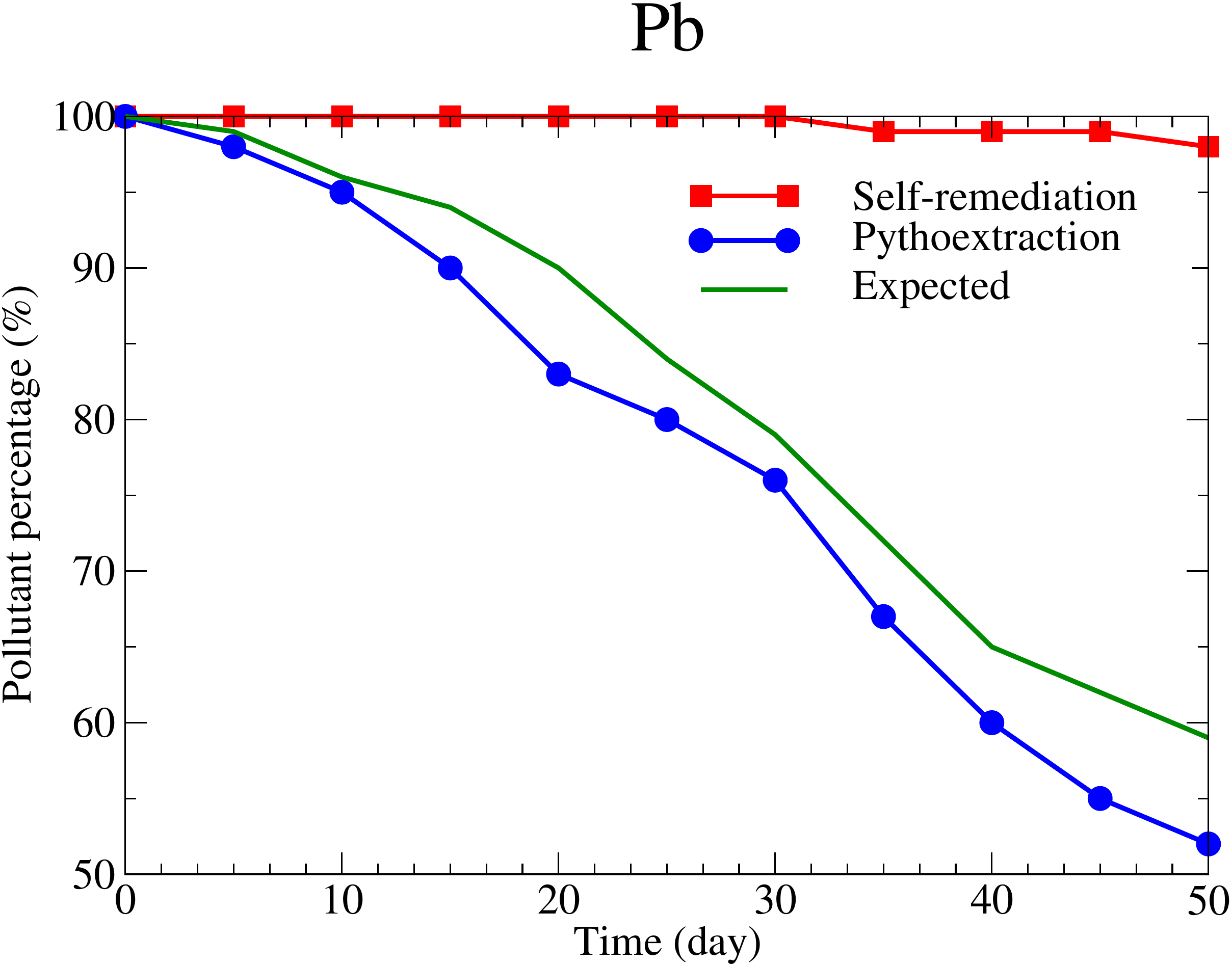}
	\caption{The remaining percentage of Pb after degradation in the shallow layer of the soil using phytoextraction technique in 50 days.}
	\label{Fig:Pb_Deg}
\end{figure}

Figure \ref{Fig:Pb_Deg} shows that the phytoextraction technique could degrade a heavy metal pollutant like Pb up to 50$\%$ after 50 days. As mentioned before, Brassica juncea consumes lead (II,IV) oxide, and after breaking its molecular structure shown in Figure \ref{Fig:3c_Lead}, transforms Pb into biomass. The degradation rate was higher after day 30 (Figure \ref{Fig:Pb_Deg}) which was due to the first harvesting of the mustards. This phenomenon proved that the full removal can be achieved where the continuous harvesting is established \citep{chibuike2014heavy}. Also as expected, the soil was not able to remediate itself from Pb contamination. 

\section{EFFECTS OF CEMENT COMPONENTS ON BIOREMEDIATION}
The presence of cement contaminations is inevitable in a concrete production site \citep{eslami_Alizadeh_Omidifard_2013}. Such a contamination could exist in three forms: (i) cement dusts which settle as sediment eventually but they are not able to complete the hydration process due to their moisture absorption from air; (ii) cement particles that have been inserted into the shallow layers of soil and eventually hydrate by absorbing the soil’s moisture; and (iii) the hydrated cement from the wastes of the concrete production \citep{khonsari2018fibrous,khonsari2010effects,eslami_2013}. Regarding the last one, other contaminations can be inserted especially when the chemical admixtures such as superplastisizers are utilized. Such an admixture contains big molecules, such as Lactates and Citrates, with a complex structure. On the other hand, in most of the concrete production sites in Iran, the polluting activities are positioned in the outdoor areas with a poor soil pre-treatment. So, the pollution sourced from the equipment activities directly drops to the soil surface. 

Regardless of the source and form of the cement contamination, the accumulation of cementitious components could harm the self-remediation of the surficial soil layers and consequently, result in a dangerous level of contamination. Regarding the deep layers of soil, the infiltration of cementitious components could result in formation of a dense structure by filling the soil pores (which supply oxygen to the biological processes) and consequently, could decrease the degradation of soil. The cement components can also affect the permeability of the soil especially in the deep layers. When the cement components penetrate into the soil, the permeability of the soil will decrease. This in turn results in decreasing the efficiency of the biological processes caused by the bioventing-biosparging technique due to the difficulty of providing the nutrient- and oxygen-rich air supplement. The aforementioned effects of the cement components cause a lower level of degradation rate in the deep layers of soil. As Figures \ref{Fig:Ethylbanzene_Deg} and \ref{Fig:Pyrene_Deg} show, the expected degradation was more than the observed values for both Ethylbenzene and Pyrene. The existence of the cementitious materials decreased the self-remediation rate of the soil, since both Ethylbenzene and Pyrene were degraded by 5$\%$ after 50 days of operation.

Although the cementitious materials negatively affect the deeper layers of the soil, their existence in a proper amount could be beneficial to the biological process. The first advantage of the cementitious components for the surficial soil is that they increase the temperature of the soil medium. The heat produced by the hydration process can increase absorption of oxygen by plant’s rhizosphere. Second, the existence of the cementitious materials in the shallow layers results in a proper stabilization and consequently, halts the infiltration of the heavy metals like Pb. This phenomenon increases the concentration of heavy metals and, as the degradation rate of a pollutant is generally proportional to the concentration, a higher degradation rate can be obtained at the shallow layer. In addition, the nutrients provided either by the chemical fertilizers (related to the previous usage of the site as an agricultural land) or by acid rains containing sulfate and nitrate which commonly fall in the region can more efficiently propagate throughout a well-stabilized soil. As Figure \ref{Fig:Pb_Deg} shows, the observation degradation level of Pb was measured to be higher than the expected curve. Note that the self-remediation of soil is not expected when the heavy metals contamination is concerned.  

\section{CONCLUSION}
In this study, the bioremediation of deep and shallow layers of the soil was presented using a case study of a construction site in Iran. The effects of the cementitious materials on the performance of the bioremediation techniques were investigated. A combined bioventing-biosparging technique was used to remediate Ethylbenzene (as a BTEX pollutant) and Pyrene (as a PAH pollutant) in the deep layers of soil. The presence of the cement components resulted in a lower rate of degradation for pollutants. This phenomenon is due to the decrease in the soil permeability and the increase in the soil density as a result of cement infiltration. The phytoextraction was used to treat Pb-polluted soil in the shallow layers of soil. The rate of degradation was increased due to the stabilization of the surficial soil by the cement components. The results of this study indicated that the proper control and monitoring of the cement component concentration is required to achieve the optimum performance of the bioremediation techniques either for the deep or the shallow layers of soil.
\bibliographystyle{plainnat}
\bibliography{Master_References/Books,Master_References/Master_References}

\begin{thebibliography}{39}
\providecommand{\natexlab}[1]{#1}
\providecommand{\url}[1]{\texttt{#1}}
\expandafter\ifx\csname urlstyle\endcsname\relax
  \providecommand{\doi}[1]{doi: #1}\else
  \providecommand{\doi}{doi: \begingroup \urlstyle{rm}\Url}\fi

\bibitem[Adams et~al.(2015)Adams, Fufeyin, Okoro, and
  Ehinomen]{adams2015bioremediation}
G.~O. Adams, P.~T. Fufeyin, S.~E. Okoro, and I.~Ehinomen.
\newblock {Bioremediation, biostimulation and bioaugmention: a review}.
\newblock \emph{International Journal of Environmental Bioremediation \&
  Biodegradation}, 3:\penalty0 28--39, 2015.

\bibitem[Azubuike et~al.(2016)Azubuike, Chikere, and
  Okpokwasili]{azubuike2016bioremediation}
C.~C. Azubuike, C.~B. Chikere, and G.~C. Okpokwasili.
\newblock {Bioremediation techniques--classification based on site of
  application: principles, advantages, limitations and prospects}.
\newblock \emph{World Journal of Microbiology and Biotechnology}, 32:\penalty0
  180, 2016.

\bibitem[Boopathy(2000)]{boopathy2000factors}
R.~Boopathy.
\newblock {Factors limiting bioremediation technologies}.
\newblock \emph{Bioresource technology}, 74\penalty0 (1):\penalty0 63--67,
  2000.

\bibitem[Cachada et~al.(2018)Cachada, Rocha-Santos, and
  Duarte]{cachada2018soil}
A.~Cachada, T.~Rocha-Santos, and A.~C. Duarte.
\newblock {Soil and Pollution: An Introduction to the Main Issues}.
\newblock In \emph{Soil Pollution}, pages 1--28. Elsevier, 2018.

\bibitem[Cai et~al.(2008)Cai, Mo, Wu, Katsoyiannis, and Zeng]{cai2008status}
Q.~Y. Cai, C.~H. Mo, Q.~T. Wu, A.~Katsoyiannis, and Q.~Y. Zeng.
\newblock {The status of soil contamination by semivolatile organic chemicals
  (SVOCs) in China: a review}.
\newblock \emph{Science of the Total Environment}, 389:\penalty0 209--224,
  2008.

\bibitem[Cao et~al.(2008)Cao, Jiang, Zeng, Du, Tan, and
  Liu]{cao2008trichoderma}
L.~Cao, M.~Jiang, Z.~Zeng, A.~Du, H.~Tan, and Y.~Liu.
\newblock {Trichoderma atroviride F6 improves phytoextraction efficiency of
  mustard (Brassica juncea (L.) Coss. var. foliosa Bailey) in Cd, Ni
  contaminated soils}.
\newblock \emph{Chemosphere}, 71\penalty0 (9):\penalty0 1769--1773, 2008.

\bibitem[Carr{\'e} et~al.(2017)Carr{\'e}, Caudeville, Bonnard, Bert, Boucard,
  and Ramel]{carre2017soil}
F.~Carr{\'e}, J.~Caudeville, R.~Bonnard, V.~Bert, P.~Boucard, and M.~Ramel.
\newblock {Soil contamination and human health: A major challenge for global
  soil security}.
\newblock In \emph{Global Soil Security}, pages 275--295. Springer, 2017.

\bibitem[Chen et~al.(2015)Chen, Xu, Zeng, Yang, Huang, and
  Zhang]{chen2015bioremediation}
M.~Chen, P.~Xu, G.~Zeng, C.~Yang, D.~Huang, and J.~Zhang.
\newblock {Bioremediation of soils contaminated with polycyclic aromatic
  hydrocarbons, petroleum, pesticides, chlorophenols and heavy metals by
  composting: applications, microbes and future research needs}.
\newblock \emph{Biotechnology Advances}, 33:\penalty0 745--755, 2015.

\bibitem[Chibuike and Obiora(2014)]{chibuike2014heavy}
G.~U. Chibuike and S.~C. Obiora.
\newblock {Heavy metal polluted soils: effect on plants and bioremediation
  methods}.
\newblock \emph{Applied and Environmental Soil Science}, 2014, 2014.

\bibitem[Dixon(1996)]{dixon1996bioremediation}
B.~Dixon.
\newblock Bioremediation is here to stay, 1996.

\bibitem[Dong et~al.(2001)Dong, Cui, and Liu]{dong2001instances}
W.~Q.~Y. Dong, Y.~Cui, and X.~Liu.
\newblock {Instances of soil and crop heavy metal contamination in China}.
\newblock \emph{Soil and Sediment Contamination}, 10:\penalty0 497--510, 2001.

\bibitem[EPA(2006)]{EPAExsituinsitu2006}
EPA.
\newblock {Engineering Issue: In Situ and Ex Situ Biodegradation Technologies
  for Remediation of Contaminated Sites}.
\newblock 62:\penalty0 6--15, 2006.

\bibitem[Erdogan and Karaca(2011)]{erdogan2011bioremediation}
E.~E. Erdogan and A.~Karaca.
\newblock {Bioremediation of crude oil polluted soils}.
\newblock \emph{Asian Journal of Biotechnology}, 3\penalty0 (3):\penalty0
  206--213, 2011.

\bibitem[Eslami(2013)]{eslami_2013}
E.~Eslami.
\newblock {The use of rice husk ash as a cementitious material: physical and
  economic analysis in Mazandaran Province, Iran}.
\newblock In \emph{Proceedings of the 2nd international conference on cement
  industry, energy and environment (CIEE), Tehran, Iran}, November 2013.

\bibitem[Eslami et~al.(2013)Eslami, Alizadeh-Javaheri, and
  Omidifard]{eslami_Alizadeh_Omidifard_2013}
E.~Eslami, M.~Alizadeh-Javaheri, and A.~H. Omidifard.
\newblock {Environmental pollution in construction sites: the case study of
  concrete element manufacturing site in Iran}.
\newblock In \emph{Proceedings of the 2nd international conference on cement
  industry, energy and environment (CIEE), Tehran, Iran}, November 2013.

\bibitem[Ferrarese et~al.(2008)Ferrarese, Andreottola, and
  Oprea]{ferrarese2008remediation}
E.~Ferrarese, G.~Andreottola, and I.~A. Oprea.
\newblock {Remediation of PAH-contaminated sediments by chemical oxidation}.
\newblock \emph{Journal of Hazardous Materials}, 152:\penalty0 128--139, 2008.

\bibitem[Joodat et~al.(2018)Joodat, Nakshatrala, and
  Ballarini]{Nakshatrala_Joodat_Ballarini_P2}
S.~H.~S. Joodat, K.~B. Nakshatrala, and R.~Ballarini.
\newblock {Modeling flow in porous media with double porosity/permeability:~A
  stabilized mixed formulation, error analysis, and numerical solutions.}
\newblock \emph{Computer Methods in Applied Mechanics and Engineering},
  337:\penalty0 632--676, 2018.

\bibitem[Joshaghani et~al.(2018)Joshaghani, Joodat, and
  Nakshatrala]{joshaghani_Joodat_Nakshatrala_2018}
M.~S. Joshaghani, S.~H.~S. Joodat, and K.~B. Nakshatrala.
\newblock {A stabilized mixed discontinuous Galerkin formulation for double
  porosity/permeability model}.
\newblock \emph{arXiv preprint arXiv:1805.01389}, 2018.

\bibitem[Khan et~al.(2004)Khan, Husain, and Hejazi]{khan2004overview}
F.~I. Khan, T.~Husain, and R.~Hejazi.
\newblock {An overview and analysis of site remediation technologies}.
\newblock \emph{Journal of environmental management}, 71:\penalty0 95--122,
  2004.

\bibitem[Khan et~al.(2017)Khan, Khan, Khan, and Alam]{khan2017soil}
M.~A. Khan, S.~Khan, A.~Khan, and M.~Alam.
\newblock {Soil contamination with cadmium, consequences and remediation using
  organic amendments}.
\newblock \emph{Science of the Total Environment}, 601:\penalty0 1591--1605,
  2017.

\bibitem[Khodaverdi et~al.(2009)Khodaverdi, Faghih, and
  Eslami]{khodaverdi2008fuzzy}
K.~Khodaverdi, A.~Faghih, and E.~Eslami.
\newblock {A Fuzzy Analytic Network Process Approach to Evaluate Concrete Waste
  Management Options}.
\newblock In \emph{Proceedings ot the 10th International Symposium on the
  Analytic Hierarchy Process (ISAHP), Pittsburgh, PA, USA}, August 2009.

\bibitem[Khonsari et~al.(2010)Khonsari, Eslami, and
  Anvari]{khonsari2010effects}
S.~V. Khonsari, E.~Eslami, and A.~Anvari.
\newblock {Effects of expanded perlite aggregate (EPA) on the mechanical
  behavior of lightweight concrete}.
\newblock In \emph{Proceedings of the 7th International Conference on Fracture
  and Mechanics of Concrete \& Concrete Structure (FraMCoS-7), Jeju, Korea},
  pages 1354--1361, 2010.

\bibitem[Khonsari et~al.(2018)Khonsari, Eslami, and
  Anvari]{khonsari2018fibrous}
S.~V. Khonsari, E.~Eslami, and A.~Anvari.
\newblock {Fibrous and non-fibrous Perlite concretes--experimental and SEM
  studies}.
\newblock \emph{European Journal of Environmental and Civil Engineering},
  22:\penalty0 138--164, 2018.

\bibitem[Koshlaf and Ball(2017)]{koshlaf2017soil}
E.~Koshlaf and A.~S. Ball.
\newblock {Soil bioremediation approaches for petroleum hydrocarbon polluted
  environments}.
\newblock \emph{AIMS Microbiol}, 3:\penalty0 25--49, 2017.

\bibitem[Krishna and Govil(2007)]{krishna2007soil}
A.~K. Krishna and P.~K. Govil.
\newblock {Soil contamination due to heavy metals from an industrial area of
  Surat, Gujarat, Western India}.
\newblock \emph{Environmental monitoring and assessment}, 124:\penalty0
  263--275, 2007.

\bibitem[Lu et~al.(2015)Lu, Song, Wang, Liu, Meng, Sweetman, Jenkins, Ferrier,
  Li, and Luo]{lu2015impacts}
Y.~Lu, S.~Song, R.~Wang, Z.~Liu, J.~Meng, A.~J. Sweetman, A.~Jenkins, R.~C.
  Ferrier, H.~Li, and W.~Luo.
\newblock {Impacts of soil and water pollution on food safety and health risks
  in China}.
\newblock \emph{Environment international}, 77:\penalty0 5--15, 2015.

\bibitem[Megharaj et~al.(2011)Megharaj, Ramakrishnan, Venkateswarlu,
  Sethunathan, and Naidu]{megharaj2011bioremediation}
M.~Megharaj, B.~Ramakrishnan, K.~Venkateswarlu, N.~Sethunathan, and R.~Naidu.
\newblock {Bioremediation approaches for organic pollutants: a critical
  perspective}.
\newblock \emph{Environment international}, 37\penalty0 (8):\penalty0
  1362--1375, 2011.

\bibitem[Mohan et~al.(2006)Mohan, Kisa, Ohkuma, Kanaly, and
  Shimizu]{mohan2006bioremediation}
S.~V. Mohan, T.~Kisa, T.~Ohkuma, R.~A. Kanaly, and Y.~Shimizu.
\newblock {Bioremediation technologies for treatment of PAH-contaminated soil
  and strategies to enhance process efficiency}.
\newblock \emph{Reviews in Environmental Science and Bio/Technology},
  5\penalty0 (4):\penalty0 347--374, 2006.

\bibitem[Nakshatrala et~al.(2018)Nakshatrala, Joodat, and
  Ballarini]{Nakshatrala_Joodat_Ballarini_a_2018}
K.~B. Nakshatrala, S.~H.~S. Joodat, and R.~Ballarini.
\newblock {Modeling flow in porous media with double
  porosity/permeability:~Mathematical model, properties, and analytical
  solutions.}
\newblock \emph{Journal of Applied Mechanics}, 85:\penalty0 081009, 2018.

\bibitem[Natsheh et~al.(2013)Natsheh, Abu-Khalaf, Sayara, Khayat, and
  Salman]{natsheh2013multivariate}
B.~Natsheh, N.~Abu-Khalaf, T.~Sayara, S.~Khayat, and M.~Salman.
\newblock {Multivariate data analysis for bioremediation of con-taminated soil
  through Interactions between heavy met-als, microbes and plants}.
\newblock \emph{Palestine Technical University Research Journal}, 1\penalty0
  (1):\penalty0 21--28, 2013.

\bibitem[Pant et~al.(2016)Pant, Bharadwaj, Wahi, Nehra, Gupta, and
  Bhatia]{pant2016bioremediation}
G.~Pant, A.~Bharadwaj, N.~Wahi, N.~Nehra, M.~Gupta, and A.~K. Bhatia.
\newblock {Bioremediation: An Eco-friendly Approach for Treating Pesticides.}
\newblock \emph{Advances in Bioresearch}, 7:\penalty0 200--206, 2016.

\bibitem[Sayara et~al.(2011)Sayara, Borr{\`a}s, Caminal, Sarr{\`a}, and
  S{\'a}nchez]{sayara2011bioremediation}
T.~Sayara, E.~Borr{\`a}s, G.~Caminal, M.~Sarr{\`a}, and A.~S{\'a}nchez.
\newblock {Bioremediation of PAHs-contaminated soil through composting:
  influence of bioaugmentation and biostimulation on contaminant
  biodegradation}.
\newblock \emph{International Biodeterioration \& Biodegradation}, 65\penalty0
  (6):\penalty0 859--865, 2011.

\bibitem[Shiomi(2013)]{shiomi2013novel}
N.~Shiomi.
\newblock {A Novel Bioremediation Method for Shallow Layers of Soil Polluted by
  Pesticides}.
\newblock \emph{Edited by Yogesh B. Patil and Prakash Rao, Applied
  Bioremediation, Active and Passive Approaches}, pages 287--306, 2013.

\bibitem[Shukla et~al.(2010)Shukla, Singh, and
  Sharma]{shukla2010bioremediation}
K.~P. Shukla, N.~K. Singh, and S.~Sharma.
\newblock {Bioremediation: developments, current practices and perspectives}.
\newblock \emph{Genetic Engineering and Biotechnology Journal}, 3\penalty0
  (8):\penalty0 1--20, 2010.

\bibitem[Strong and Burgess(2008)]{strong2008treatment}
P.~J. Strong and J.~E. Burgess.
\newblock {Treatment methods for wine-related and distillery wastewaters: a
  review}.
\newblock \emph{Bioremediation Journal}, 12:\penalty0 70--87, 2008.

\bibitem[Tiller(1992)]{tiller1992urban}
K.~G. Tiller.
\newblock {Urban soil contamination in Australia}.
\newblock \emph{Soil Research}, 30:\penalty0 937--957, 1992.

\bibitem[Vidali(2001)]{vidali2001bioremediation}
M.~Vidali.
\newblock {Bioremediation. an overview}.
\newblock \emph{Pure and Applied Chemistry}, 73\penalty0 (7):\penalty0
  1163--1172, 2001.

\bibitem[Wang et~al.(2003)Wang, Cui, Liu, Dong, and Christie]{wang2003soil}
Q.~R. Wang, Y.~S. Cui, X.~M. Liu, Y.~T. Dong, and P.~Christie.
\newblock {Soil contamination and plant uptake of heavy metals at polluted
  sites in China}.
\newblock \emph{Journal of Environmental Science and Health, Part A},
  38:\penalty0 823--838, 2003.

\bibitem[Wilcke et~al.(1998)Wilcke, M{\"u}ller, Kanchanakool, and
  Zech]{wilcke1998urban}
W.~Wilcke, S.~M{\"u}ller, N.~Kanchanakool, and W.~Zech.
\newblock {Urban soil contamination in Bangkok: heavy metal and aluminium
  partitioning in topsoils}.
\newblock \emph{Geoderma}, 86:\penalty0 211--228, 1998.

\end{thebibliography}
\end{document}